\documentclass[pre, amsmath, amssymb, superscriptaddress, twocolumn]{revtex4}

\usepackage{graphicx}
\usepackage{dcolumn}
\usepackage{bm}
\usepackage{braket}
\usepackage{xcolor}

\usepackage{amsmath}
\usepackage{SIunits} 

\newcommand{\beq}{\begin{equation}}
\newcommand{\eeq}{\end{equation}}

\begin{document}


\title{Does blood type affect the COVID-19 infection pattern?}



\author{Mattia Miotto\footnote{Corresponding author: mattia.miotto@roma1.infn.it}}
\affiliation{Department of Physics, Sapienza University, Piazzale Aldo Moro 5, 00185, Rome, Italy}
\affiliation{Center for Life Nanoscience, Istituto Italiano di Tecnologia, Viale Regina Elena 291,  00161, Rome, Italy}

\author{Lorenzo Di Rienzo}
\affiliation{Center for Life Nanoscience, Istituto Italiano di Tecnologia, Viale Regina Elena 291,  00161, Rome, Italy}

\author{Giorgio Gosti}
\affiliation{Center for Life Nanoscience, Istituto Italiano di Tecnologia, Viale Regina Elena 291,  00161, Rome, Italy}

\author{Edoardo Milanetti}
\affiliation{Department of Physics, Sapienza University, Piazzale Aldo Moro 5, 00185, Rome, Italy}
\affiliation{Center for Life Nanoscience, Istituto Italiano di Tecnologia, Viale Regina Elena 291,  00161, Rome, Italy}

\author{Giancarlo Ruocco}
\affiliation{Center for Life Nanoscience, Istituto Italiano di Tecnologia, Viale Regina Elena 291,  00161, Rome, Italy}
\affiliation{Department of Physics, Sapienza University, Piazzale Aldo Moro 5, 00185, Rome, Italy}


\begin{abstract}
Among the many aspects that characterize the COVID-19 pandemic, two seem particularly challenging to understand:
{\it i}) the great geographical differences 
in the degree of virus contagiousness
and lethality which were found
in the different phases of the epidemic progression, and,
{\it ii}) the potential role of the infected people's blood type in both the virus infectivity and the progression of the disease. A recent hypothesis could shed some light on both aspects. Specifically, it has been proposed that in the subject-to-subject transfer SARS-CoV-2 conserves on its capsid the erythrocytes' antigens of the source subject. Thus these conserved antigens can potentially cause an immune reaction in a receiving subject that has previously acquired specific antibodies for the source subject antigens. This hypothesis implies a blood type-dependent infection rate. The strong geographical dependence of the blood type distribution could be, therefore, one of the factors at the origin of the observed heterogeneity in the epidemics spread. Here, we present an epidemiological deterministic model where the infection rules based on blood types are taken into account and compare our model outcomes with the exiting worldwide infection progression data. We found an overall good agreement, which strengthens the hypothesis that blood types do play a role in the COVID-19 infection.
\end{abstract}

\flushbottom
\maketitle
%
%
\thispagestyle{empty}


\section*{Introduction}
The new infectious coronavirus disease 2019, called COVID-19, began to spread from China in December 2019 \cite{wang2020novel}. The most evident COVID-19 symptoms are pneumonia and respiratory failure,  which reiterate the symptoms reported in the SARS (Severe Acute Respiratory Syndrome) epidemic of 2003 \cite{huang2020clinical,zhu2020novel}. 
The first cluster to clearly show these symptoms were patients from Wuhan, People’s Republic of China (WMHC) \cite{huang2020clinical}. In early January 2020, scientists at the National Institute of Viral Disease Control and Prevention (IVDC) isolated the new virus for the first time from patients in Wuhan and found it to be a novel $\beta$-genus coronavirus, which has been named SARS-CoV-2~\cite{Team2020}.
Currently, the outbreak has rapidly spread in many other countries. Hence, on 11 March 2020, the World Health Organization declared it a pandemic \cite{WHO, walls2020structure}.


Understanding the transmission dynamics of this infection plays a key role in assessing the diffusion potential that may be sustained in the future.
In this context, models and simulations represent a powerful tool, which can be useful to study and monitor human and animal viral infections \cite{anderson1979population, martinez2011novel}. These tools have become fundamental, especially during this pandemic, to evaluate the trade-off between cost and effectiveness of various social distancing strategies, and to enable policymakers to make the best decisions in the interest of the public health~\cite{Prem2020}.
Nonetheless, each disease is characterized by its specific biological rules, therefore it is essential to consider them in a mathematical model to describe real situations.
As the virus spreads across the world the pandemic has presented a similar pattern, in all the countries that recorded a significant number of infections. The pattern is made up of a first phase that is characterized by an exponential increase of infections and a later phase in which the implementation of social distancing measures reduces the spread of the disease to a sub-exponential growth, which generally is followed by a gradual decrease of daily infections. Eventually, the gradual decrease in the number of daily infections becomes smaller than the daily recovered number, thus the number of the total infected starts decreasing.
Even if this general pattern has been reproduced around the world, the spread of the virus showed important local differences, mostly in the rate of the initial exponential spread.  Given the complex nature of this historic event, it is extremely difficult to understand if these patterns are the consequence of geographical inhomogeneities or if these are spurious correlations which are caused by the singularity of the observed event. 
Indeed, some works underlined as in the early stage of the epidemics the virus showed a geographical pattern with most of the infection localized in temperate regions characterized by specific characteristics of temperature and humidity~\cite{Araujo2020, OReilly2020, Sajadi2020, Scafetta2020}.
Furthermore, in the specific context of Italy, a non-negligible difference in the rate of infection and mortality between the Northern and Southern regions was reported. But most analyses considered these differences as the effect of the implementation of social distancing rules before that the disease spread in South Italy and not of climate variables \cite{Sebastiani2020}.
Other potential co-morbidities that may explain local patterns are hypertension, obesity, and age distribution, which are known to display heterogeneous local distributions~\cite{Kershaw2010, Samouda2018}. 
Also, the local history of past infections of different coronaviruses could contribute to the observed heterogeneity, due to cross-reactivity immunity effects~\cite{Grifoni2020, Sette2020}.

In particular, blood groups were recognized to influence susceptibility to certain viruses, including SARS-CoV-1~\cite{cheng2005abo}  and norovirus~\cite{hutson2002norwalk}.  Blood group A and B glycosyltransferases also affect glycosylation in a large number of cell types, including epithelial cells in the respiratory tract and shed viral particles~\cite{yamamoto2012abo}.

\begin{figure}[t]
\centering
\includegraphics[width=\linewidth]{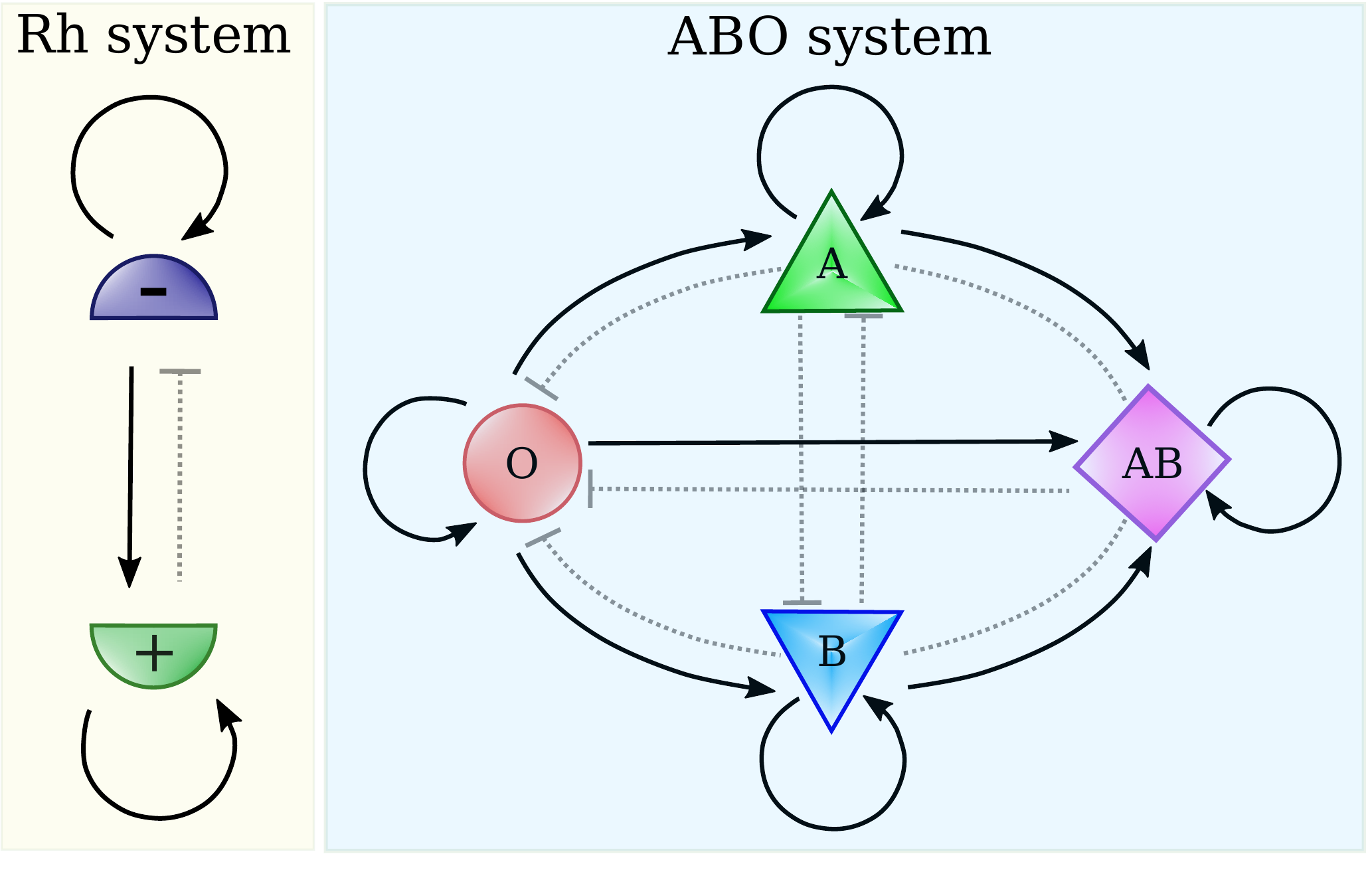}
\caption{Scheme of the possible infection rules according to Rh blood types and ABO ones. Full line connection indicated the possibility of infection, $W_{ij}=1$,  while dotted connection its impossibility. In the Rh infection system, individuals of the same group can infect each other, people with the Rh- group can transmit the infection to people with Rh+ one, but not \textit{viceversa}. The scenario of the ABO system is similar although richer. For instance, $0$ type can infect the $A$ type, while the opposite is not possible. 
} 
\label{fig6} 
\end{figure}

Recently, Zhao \emph{et al.}~\cite{zhao2020relationship} found that ABO blood groups
presented a different risk to contract COVID-19 as a result of being exposed to SARS-CoV-2.
Previously, for the similar coronavirus SARS-CoV responsible for SARS,
Guillon \emph{et al.}~\cite{guillon2008inhibition} showed experimentally that for SARS-CoV synthesized by cells that expressed the A histo-blood group antigen, the interaction between S protein and its membrane receptor, ACE2, could be blocked by anti-A blood group antibodies.
Starting from these experimental results regarding the SARS-CoV spike, Breiman \emph{et al.}~\cite{breiman2020harnessing} extended the hypothesis to the new SARS-CoV-2, suggesting that the different susceptibility of individuals with different ABO blood groups may have the same explanation. 
This new hypothesis, based on the infection rules schematically illustrated in Figure~\ref{fig6},  may explain a part of the variability in the infection contagiousness among the countries of the world \cite{breiman2020harnessing}.

Here, we discuss the spread of the epidemic from a mathematical modeling perspective, taking into account the influence of the different sub-populations divided according to blood types, in the different geographical areas. The main purpose of this work is to verify whether the hypothesized blood group based infection rules can be consistent with the acquired country-level data on blood group distribution and infection diffusion dynamics. 
In the last decades, the modeling of epidemic diffusion behaviors has rapidly developed and found wide application~\cite{vespignani2012modelling}. Many recent papers  described different aspects of the COVID-19 disease evolution through  mathematical models~\cite{ivorra2020mathematical, roosa2020real, kucharski2020early, yang2020modified, lin2020conceptual,Boudrioua2020,2003.14391,tang2020estimation}.
We propose here a generalization of the widely-used Susceptible-Infected-Recovered model (SIR), a simple compartmental model, where the population is divided into three sets. ``Susceptible" are healthy individuals who can be infected by other individuals. ``Infected" individuals are those who contracted the virus at time t, while  ``Recovered” consists of all the immune individuals, therefore they cannot be infected again. 
Each of these compartments is time-dependent, owing to the progressive nature of the disease.
Since the SIR model is a deterministic system, we can model the epidemic process using ordinary differential equations.
Consequently, the evolution in time of each compartment state is completely determined by the initial conditions along with the differential equations~\cite{Menon2020}.
Here, we modified the SIR model to take into account the proposed blood type infection rules to test the working hypothesis against real data.
     
\section*{Model}

In this section we first briefly consider the standard SIR model for the time evolution of the fraction of susceptible  ($x(t)$), infected  ($y(t)$), and recovered ($z(t)$) people in a population of size $N$, recalling the main features of the model and its solution~\cite{Harko2014}. 

We then analyze in detail the generalization of the SIR model to a more complex pattern of infection, as the one described in the previous section.
A more complete discussion about the mathematical details of this treatment can be found in Supporting Information.

\subsection*{Standard SIR model}

The standard SIR model reads
\begin{eqnarray}
& \frac {dx(t)}{dt} &  = -\beta \;x(t) \; y(t) 
\label{SIR1-xt}\\
& \frac {dy(t)}{dt}  & = \beta \; x(t) \; y(t) - \gamma \; y(t) 
\label{SIR1-yt}\\
& \frac {dz(t)}{dt}  & = \gamma \; y(t)
\label{SIR1-zt}
\end{eqnarray}
where the three functions satisfy the relation 
\begin{equation}
\label{SumRule}
x(t)+y(t)+z(t)=1
\end{equation}

and the parameter $\beta$ and $\gamma$ represent the infection and recovery rate respectively.

We introduce the parameter $\rho  \doteq \gamma / \beta$ and the re-scaled time $\tau \doteq \beta t$.  We thus chose the simple and realistic case of a few infected persons at time zero, and all the remaining population susceptible of infection:

\begin{eqnarray}
& x(0) &  = x_o \nonumber \\
& y(0) & = y_o = 1-x_o \label{zero} \\
& z(0) & = 0. \nonumber 
\end{eqnarray}

The solution of the standard SIR model that obey these initial conditions can be written as:

\begin{eqnarray}
& x(\tau) &=x_o \; e^{-z(\tau)/\rho} 
\label{stdSIRsolution_x} \\[8pt]  
& y(\tau) &=1-z(\tau)-x_o \; e^{-z(\tau)/\rho} 
\label{stdSIRsolution_y} \\
& \tau &=\frac{1}{\rho} \; \int_0^{z(\tau)}  \frac{d\zeta}{1-\zeta-x_o  e^{-\zeta/\rho} }.
\label{stdSIRsolution_tau}
\end{eqnarray}

Further details on the derivation of this solution are reported in the Supporting Information.

\subsubsection*{Short time expansion}\label{sec:stexp}

It is well know that the initial stage of the infection is well represented by an exponential growth. It is therefore useful to perform the expansion of the solutions (\ref{stdSIRsolution_x}), (\ref{stdSIRsolution_y}) and (\ref{stdSIRsolution_tau}) for $\tau \rightarrow 0$.

In the small $\tau$ limit, the expression for $\tau(z)$ given by  Eq. (\ref{stdSIRsolution_tau}), becomes:

\begin{equation}
\label{exp_tau}
\tau(z) \approx \tau|_{z=0} + \frac{d\tau}{dz} \Big |_{z=0} z = \frac{1}{\rho}   \frac{z}{(1-x_o)},
\nonumber
\end{equation}
thus

\begin{equation}
\label{exp_z}
z(\tau) \approx \rho (1-x_o) \tau = \rho y_o \tau.
\end{equation}

By substituting in (\ref{stdSIRsolution_x}) and (\ref{stdSIRsolution_y}), we get

\begin{eqnarray}
& x(\tau) &  \approx x_o \; e^{-y_o \tau} \\
& y(\tau) &  \approx 1- \rho y_o \tau-x_o \; e^{-y_o \tau}.
\end{eqnarray}

The last equation, by expanding the exponential, collecting the terms linear in $\tau$ and resuming the exponential, becomes:
\begin{equation}
\label{exp_yy}
y(\tau) \approx y_o \; e^{(x_o-\rho) \tau} 
\end{equation}
which represents the desired exponential growth of the infection at short time. As at short time the fraction of infected population is very small (i.e. $y_o \ll 1$), we can safely approximate $x_o \approx 1$:
\begin{equation}
\label{exp_yyy}
y(\tau) \approx y_o \; e^{(1-\rho) \tau}.
\end{equation}

Equation~(\ref{exp_yyy}) expresses one of the key concepts of epidemic models. The number of infectious individuals grows exponentially if $\rho < 1$, or $\beta > \gamma$. Often, in the literature, the parameter controlling the level of infection growth is the ``reproduction number", $R_o$, defined as the average number of secondary infections caused by a primary case introduced in a fully susceptible population~\cite{anderson1992infectious}. $R_o$ is therefore equal to $\beta / \gamma$, which, in our notation, means $R_o=1/\rho$.

From those simple considerations arises the concept of epidemic threshold: only if $(1-\rho) >0$, thus $R_0 > 1$ (i.e. if a single infected individual generates on average more than one secondary infection), an infective agent can cause an outbreak. If $R_0 < 1$ (i.e. if a single infected individual generates less than one secondary infection), then $(1-\rho)<0$ and the initial stage of the epidemy is characterized by a \textit{decreasing} number of cases. 

\subsubsection*{Summary of standard SIR properties}

In this subsection, we recollect the main results from previous Sections and those in the Supporting Information. 
The main results are four. First, it is worth to always perform the limit $x_o \rightarrow 1$, while keeping $y_o \ne 0$ as discussed below.
This allows us to correctly capture the exponential growth that characterizes the initial phase of an epidemic.

Second,the exact solutions of the standard SIR model wit the initial conditions (\ref{zero}) are:
\begin{eqnarray*}
x(\tau)&=& e^{-z(\tau)/\rho}
\label{SUMx}\\ [8pt]  
y(\tau)&=&1-z(\tau)-e^{z(\tau)/\rho}.
\label{SUMy}\\
\tau&=&\frac{1}{\rho} \; \int_0^{z(\tau)}  \frac{d\zeta}{1-\zeta-e^{-\zeta/\rho} }
\label{SUMz}
\end{eqnarray*}

Third, the value of maximum infection is found to be:
\begin{equation*}
\label{SUM2}
y_{M}= 1-\rho+\rho \log(\rho) .
\end{equation*}


Fourth, at short time, when the exponential growth dominate the solution, the following approximations hold:
\begin{eqnarray}
x(\tau)&=& e^{-y_o \tau}
\label{SUMxE} \nonumber\\
y(\tau)&=&y_o \; e^{(1-\rho) \tau}
\label{SUMyE}\\ [2pt]  
z(\tau)&=&  \rho y_o \tau
\label{SUMzE} \nonumber
\end{eqnarray}

As last observation, it is worth to emphasize that Eq. (\ref{SUMyE}) is identical to the short time expansion of the infectivity in the SIS (Susceptible-Infected-Susceptible) model, which is a simplified version of the SIR model where individuals never acquire immunity to the infection. The latter, at short time, reads:
\begin{equation}
\label{SIS_sol_shorttime}
\dot y(\tau) = x_o y(\tau) - \rho y(\tau)
\end{equation}
whose solution coincides with Eq. (\ref{exp_yy})) or, after the $x_o=1$ approximation, with Eq. (\ref{exp_yyy})).

\subsection*{Generalized SIR model}\label{sec:sir_gen}

In this section, we present a general SIR model, which describes the evolution of the epidemic assuming that transmission of the infection depends on the blood groups of the individuals.
In particular, we will show how to derive proper descriptors that take into consideration not only transmission rules based on the AB0 group but also possible combinations of different groups. As examples, we discuss the hypothetical cases of a two-group set of rules (i.e. the Rhesus group), the ABO group, and their combination. We note that from a biological point of view, it seems improbable that the Rh system could play a role in the infection transmission however the derived framework could be applied to other kinds of glycan-based groups, like the Lewis one. 

As we said, our aim is to generalize the SIR model to the case where the population is not homogeneous, but it is composed of different sub-populations that follow specific infection rules.

In the general case of $k$ sub-populations $i=1,...,k$ , the time evolution of the variables can be written as:

\begin{eqnarray}
&  \dot x_i &  = - x_i \; \sum_{j=1}^k W_{ij} y_j 
\label{SIR-xg}\\
& \dot y_i   & = x_i \;  \; \sum_{j=1}^k W_{ij} y_j  - \rho \; y_i
\label{SIR-yg}\\
&  \dot z_i   & = \rho \; y_i
\label{SIR-zg}
\end{eqnarray}
where the matrix $\bf W$ encodes the infection rules. 

As fisrt example, we discuss the simple case of only two sub-populations that could follow the infection rules associated to the Rh$\pm$ blood type ($f_1 \equiv f_-$, $f_2 \equiv f_+$, where $f_\pm$ is the frequency of one blood type in the population). We assume the the sub-population with Rh$+$ cannot infect the Rh$-$ one. On the other hand, the Rh$-$ sub-population can infect both Rh$-$ and Rh$+$ sub-populations. The matrix $\bf W^{(2)}$ in this situation turns out to be:
\begin{equation} \bf W^{(2)} = \left( \begin{array}{cc} 1 & 0 \\ 1 & 1 \end{array} \right) \label{W2}\end{equation}

\begin{figure}[t]
\centering
\includegraphics[width=\linewidth]{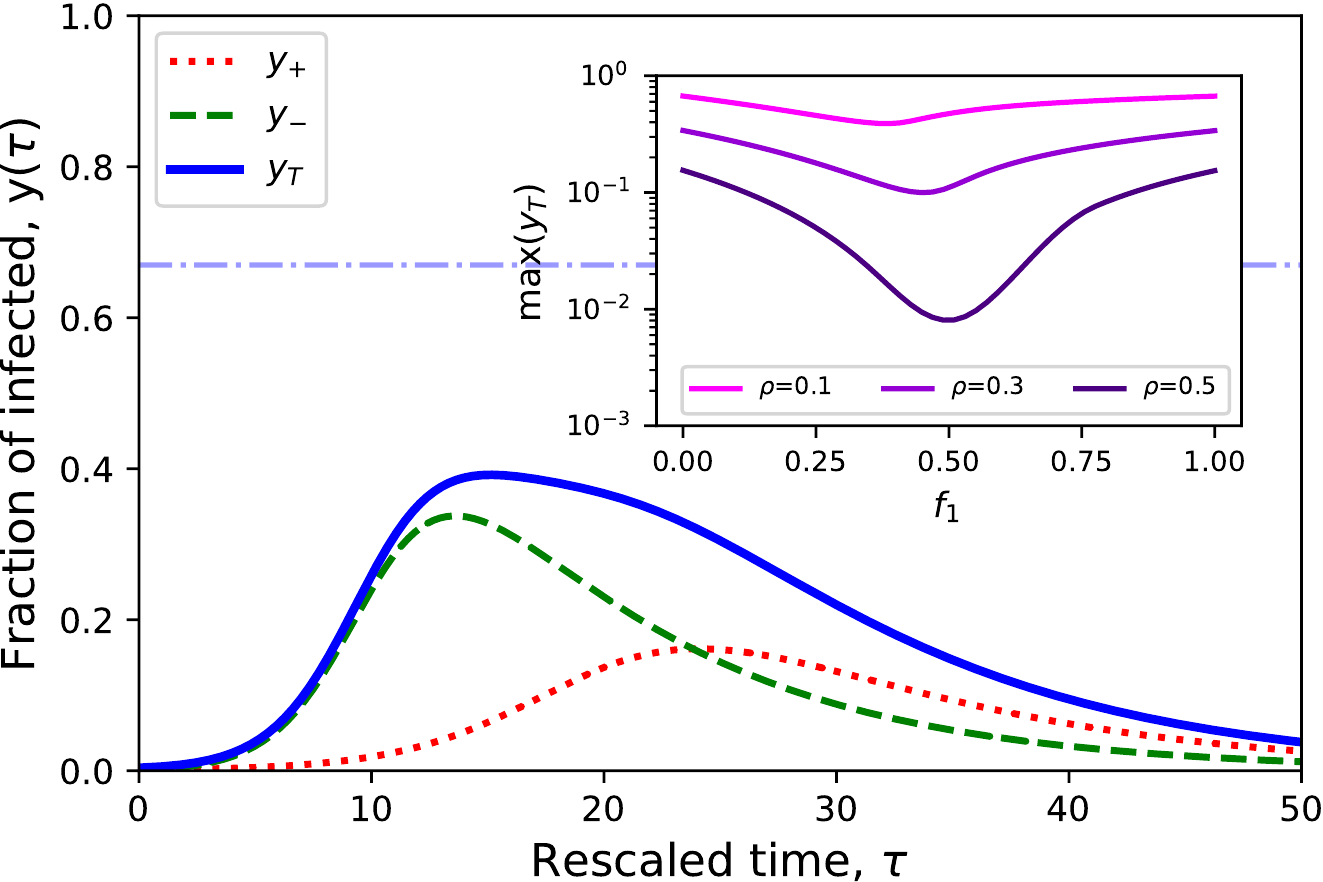}
\caption{Example of numerical solution of the generalised SIR model for two sub-populations ($k=2$) and for the infectivity matrix of Eq. (\ref{W2}). The reported example is for $\rho=0.1$, $f_1=0.4$ and $f_2=0.6$. The dashed and red lines represent the fraction of infected people in the two sub-populations, the blue line the total fraction of infected persons, the horizontal dotted line indicates the maximum value, $y_M$, of $y_T(\tau)$ in the case of all-infect-all rule. The inset report the maximum of $y_T(\tau)$ as a function of $f_1$ ($f_2=1-f_1$) for three distinct values of $\rho$.
} 
\label{fig1} 
\end{figure}

The corresponding SIR equations, to our knowledge, cannot be solved by quadrature. 
This example is discussed in some details in Supporting Information, where it is also reported the short time expansion. 
The numerical solution for this $k=2$ case is also depicted in Figure~\ref{fig1}, where as an example the fraction of infected people $y_i(\tau)$ for the two sub-populations (green dashed and red dotted lines) and the total fraction $y_T(\tau)$ of infected people (blue line) are reported as a function of the reduced time $\tau$ for the case $\rho=0.1$ and $f_1=0.4$, $f_2=0.6$. 

One can immediately notice that the maximum fraction of infected people,  $y_M$, is reduced with respect to the one-population case (horizontal dash-dotted blue line). This is a trivial consequence of the impossibility of `+' to infect `-', thus reducing the ``effective" infection rate. The dependence of the maximum infectivity on $f_1$ ($f_2=1-f_1$) is reported in the inset of Figure~\ref{fig1} for three representative values of $\rho$ ($\rho$=0.1, 0.3 and 0.5). Noteworthy, for the case $\rho$=0.5, i.e. $R_0$=2, the actual value observed in Europe, the maximum of infectivity is reduced from the all-infect-all case ($\approx$0.2) to a value more than ten times smaller ($\approx$0.01) when $f_1$ approach 0.5. The immediate consequence of this finding is that the time evolution of the epidemic strongly depends on the blood type distribution, giving a qualitative explanation of the observed high geographical variability: even a small change in $f_1/f_2$ highly affects both the infectivity maximum and the infectivity growth rate in the initial exponential growth phase. 

If we turn to consider the more biologically relevant case of the ABO types, we have four subpopulations
with the corresponding frequencies:
$f_1 \equiv f_0$, $f_2 \equiv f_A$, $f_3 \equiv f_B$, $f_4 \equiv f_{AB}$).
With the infection rules summarised in Figure~\ref{fig6} the $\bf W^{(4)}$ matrix results:
\begin{equation}  \bf W^{(4)} = \left( \begin{array}{cc}  \bf W^{(2)} & \bf 0 \\  \bf W^{(2)} &  \bf W^{(2)} \end{array} \right) \label{W4}\end{equation}

Finally, assuming that both the ABO and the Rh types would play a role, there are eight subpopulations with respective frequencies $f_{O-} \equiv f_1$, $f_{A-} \equiv f_2$, $f_{B-} \equiv f_3$, $f_{AB-} \equiv f_4$, $f_{O+} \equiv f_5$, $f_{A+} \equiv f_6$, $f_{B+} \equiv f_7$, $f_{AB+} \equiv f_8$, and the corresponding matrix $\bf W^{(8)}$ reads:
\begin{equation} \bf W^{(8)} = \left( \begin{array}{cc}  \bf W^{(4)} & \bf 0 \\  \bf W^{(4)} &  \bf W^{(4)} \end{array} \right) \label{W8} \end{equation}

The recursive structure of Eqs. (\ref{W2} - \ref{W8}) has a simple explanation, that can be better understood if we think of the ABO blood type system as the combination of two codominant and one recessive allele that form two independent systems. Let's specify if a person does have (``$+$") or does not have (``$-$") the antigen A (A$+$ or A$-$ respectively), analogously for the antigen B (B$+$ or B$-$). Then the usual blood type is: 0 $\equiv$ [A$-$,B$-$]; A $\equiv$ [A$+$,B$-$]; B $\equiv$ [A$-$,B$+$]; AB $\equiv$ [A$+$,B$+$]. With this notation it is clear that the  ``ABO" system (which follows $W^{(4)}$) is the product of the ``A" system (which follows $W^{(2)}$) and the ``B" system (which again follows $W^{(2)}$): ABO=A$\pm$ $\times$ B$\pm$. Furthermore, by multiplying the ABO system by the Rh$\pm$ system we get the ABO $\times$ Rh$\pm$ = A$\pm$ $\times$ B$\pm$ $\times$ Rh$\pm$ system, which obeys the $W^{(8)}$ infection rules. Summing up, any time it exists a set of antigens, A$_i$, $i=1..a$, that can be either present or absent, the infection rule of the A$_1$ $\times$ A$_2$ $\times$.... $\times$ A$_a$ system follows the $W^{(k)}$ infection rule with $k=2^a$.  

The set of $3k$ differential equations (\ref{SIR-xg} - \ref{SIR-zg}), together with the initial conditions:

\begin{eqnarray}
&  x_i(0) &  = f_i x_o \approx f_i
\label{SIR-x0} \nonumber\\
& y_i(0)   & = f_i y_o
\label{SIR-y0}\\
&  z_i(0)   & = 0
\label{SIR-z0} \nonumber
\end{eqnarray}
and the sum rule 
\begin{equation}
\label{sr}
\sum_{i=1}^k f_i = 1
\end{equation}
allows one to work out the short time expansion for the infected person fraction on each of the $k$ sub-populations:
\begin{equation}
\label{yi}
y_i(\tau) = y_o f_i\; e^{\pi^{(k)}_{i} \tau} \;\;\;\;\;\;\;\;\;\;\;\;\; i=1,...,k
\end{equation}
with 
\begin{equation}
\label{pi_i}
\pi^{(k)}_{i}= \sum_j W_{ij} f_j -  \rho =   ({\bf W} \cdot \bar f)_i -\rho \doteq p^{(k)}_{i} -\rho,
\end{equation}
as well as the total number of infected persons:
\begin{equation}
\label{yyyy}
y_T(\tau) = \sum_i y_i(\tau) = y_o \; e^{\pi^{(k)}_{T} \tau} \;\;\;\;\;\;\;\;\;\;\;\;\; i=1,...,k
\end{equation}
with
\begin{equation}
\label{pi}
\pi^{(k)}_{T}= \sum_i \sum_j f_i W_{ij} f_j -  \rho = \bar f  \cdot {\bf W} \cdot  \bar f -\rho \doteq p^{(k)}_{T} -\rho
\end{equation}
which implicitly define $p^{(k)}_{i}$ and $p^{(k)}_{T}$.
This simple expression for the inverse of the characteristic time of the infection at its initial stage is the sum of two terms: , $\pi^{(k)}_{T}=p^{(k)}_{T}-\rho$. 

The first one, 

\begin{equation}
\label{eq:p_k}
p^{(k)}_{T}\doteq \bar f  \cdot {\bf W} \cdot  \bar f
\end{equation}

depends only on the abundance of the sub-population ($\bar f$) and on the infection rules ($\bf W$), the second ($\rho=\gamma/\beta$) on the overall recovery and infection rates of populations. 

To study the global effect of the population composition on the progression of the infection, we concentrate on the term $p^{(k)}_{T}$ which acts as a ``susceptibility". 

It is worth to note that, for any $k$, the susceptibility is maximum when $\bar f=(0...0, 1, 0, ...0)$, $p^{(k)}_{T}=1$, i.e. when one sub-population fraction dominates,  while it minimum when the sub-populations are all of the same size: $\bar f=(1/k, 1/k ... 1/k)$. In the latter case, for the infection rules reported before, $p^{(k)}_{T}=(3/4)^{(k/2)}$. Thus the susceptibility decreases on increasing the number of sub-populations and decreases on equalizing their abundances.
Infection rules and compositions of the population are expected to shape the infection dynamics along with the usual infection and recovering rates.

In the following we will analyze available datasets of individuals infected by SARS-CoV-2, stratified by blood types. Then, we will deal with the contagion curves of a large set of countries and try to assess whether the present model could represent the real outcome of the COVID-19 pandemics.
Specifically, we will first compare data collected in the Chinese regions of Wuhan and Shenzhen reported in~\cite{zhao2020relationship}, in Denmark~\cite{Barnkob2020}, in the Ankara region (Turkey) \cite{32496734}, in New York City (NY, USA) \cite{Zietz2020}, in Italy and Spain~\cite{ellinghaus2020genomewide} and USA~\cite{leaf2020abo}. This allows for a direct comparison between our model predictions and real data. 
Then we move to analyze the early stages of the infection in different countries at a worldwide level. To take into account differences like lifestyle, climate, geographic location, and other factors that likely influence the epidemics rates as well, we identified four major areas, i.e. Europe, Asia, Africa, and South America.     

\section*{Materials and Methods}

\subsection*{Cases where infection data are stratified by blood type} 


To test the hypothesis that blood groups could impact the COVID-19 infection spread, we collected data from clinical observations of individuals found positive for SARS-CoV-2 for which also ABO blood group information was available. In Table~\ref{tab1}, the ABO frequencies of positive patients ($d_i$) and those of  control populations ($f_i$) are reported.
The first four sets of data were collected in three Wuhan hospitals and one Shenzhen hospital~ \cite{zhao2020relationship, Li2020}.
The number of patients in the three sub-cases are 1775, 113,265, and 285, respectively. 
Notably,  the abundance of each ABO blood group on the local population is reported as controls, allowing us to study the impact of local ABO phenotypic heterogeneities.
Differences between infected and control frequencies were found with a higher level of statistical significance in A and O groups. 

\begin{table*}[]
\centering
\caption{Data used to determine the quantities plotted in Figure~\ref{fig_cinesi} and Figure~\ref{fig_all}: $\ln(d_i / f_i)$ and $p_T^{(4)}$ following Eq. (\ref{pi_i}), with $d_i$ being the fractions of infected having blood group $i$, $f_i$ is the fraction of population with blood group $i$ and  $p_T^{(4)}$ represent the susceptibility of the population to become infected.}
\label{tab1}
\begin{tabular}{|c|c|cccc|c|}
\hline 
{\bf Dataset} & {\bf Quantity} & {\bf 0}  & {\bf A}  & {\bf B}  & {\bf AB}& Ref.  \\
\hline 
\hline 
Wuhan Jinyintan Hospital  &  $f_i$ &0.3384 &  0.3216 &  0.2491 &  0.091  & \cite{zhao2020relationship}\\
& $d_i$ & 0.258  &  0.3775 &  0.2642 &  0.1003 & \\
\hline
\hline
Renmin Hospital of Wuhan & $f_i$ &       0.3384 &  0.3216 &  0.2491 &  0.091  & \cite{zhao2020relationship}\\
& $d_i$ &       0.2478 &  0.3982 &  0.2212 &  0.1327 & \\ 
\hline
\hline
Shenzhen Third People Hospital & $f_i$ &   0.3877 &  0.2877 &  0.2514 &  0.0732  & \cite{zhao2020relationship}\\ 
& $d_i$ &       0.2842 &  0.2877 &  0.2912 &  0.1368 & \\
\hline
\hline
Central Hospital of Wuhan & $f_i$ &       0.3384 &  0.3216 &  0.2491 &  0.091  & \cite{Li2020}\\ 
& $d_i$ &       0.2566 &  0.3925 &  0.2528 &  0.0981 & \\
\hline
\hline
Denmark & $f_i$ &       0.417  &  0.4238 &  0.1145 &  0.0447 & \cite{Barnkob2020}\\
& $d_i$ &       0.3841 &  0.4441 &  0.1209 &  0.0509 & \\
\hline
\hline
Hacettepe Hospital of Ankara & $f_i$ &       0.3725 &  0.3804 &  0.1472 &  0.0999 & \cite{32496734} \\
& $d_i$ &       0.2473 &  0.5699 &  0.1075 &  0.0753 & \\
\hline
\hline
New York Presbyterian Hospital & $f_i$ &       0.4814 &  0.3274 &  0.1491 &  0.0421 & \cite{Zietz2020}\\
& $d_i$ &       0.4575 &  0.3416 &  0.1701 &  0.0308 & \\
\hline
\hline
Italy  & $f_i$ &       0.4709 &  0.3594 &  0.1299 &  0.0398 & \cite{ellinghaus2020genomewide}\\
& $d_i$ &       0.3749 &  0.4647 &  0.1090  &  0.0515 & \\
\hline
\hline
Spain & $f_i$ &       0.4863 &  0.4189 &  0.0684 &  0.0263 &  \cite{ellinghaus2020genomewide}\\
& $d_i$ &       0.3755 &  0.4865 &  0.0916 &  0.0465 & \\
\hline
\hline
USA (White non hispanic) & $f_i$ &       0.4525 &  0.3974 &  0.1091 &  0.041  & \cite{leaf2020abo}\\
& $d_i$ & 0.3779 &  0.451  &  0.1141 &  0.057  & \\
\hline
\hline
USA (Black non hispanic) & $f_i$ &       0.502  &  0.258  &  0.197  &  0.043  & \cite{leaf2020abo}\\
&  $d_i$ &      0.4791 &  0.2713 &  0.2171 &  0.0326 & \\
\hline
\hline
USA (Asian non hispanic) & $f_i$ &       0.3976 &  0.2777 &  0.2537 &  0.0709 & \cite{leaf2020abo}\\
& $d_i$ &       0.2982 &  0.2807 &  0.3246 &  0.0965 & \\      
\hline
\hline
USA (Hispanic) & $f_i$ &       0.5650  &  0.3110  &  0.0990  &  0.0250  & \cite{leaf2020abo}\\
& $d_i$ &       0.6127 &  0.2941 &  0.0686 &  0.0245 & \\
\hline
\hline
\end{tabular}
\end{table*}

A recent work by Barnkob \textit{et al.}~\cite{Barnkob2020} retrieved ABO group information for 7422 Danish individuals found positive to SARS-CoV-2 between 27 February 2020 and 30 July 2020.  The reference ABO frequencies, $f_i$ were instead obtained from about 2 million Danish people.
In this case, statistical confidence is found for O,A, and AB groups (p-value $<0.05$), while B group is associated to a p-value of 0.091 (see~\cite{Barnkob2020}).


Further data stratified by ABO blood groups are reported in~\cite{32496734} for 186 patients with a confirmed diagnosis of COVID-19 in the region of Ankara. The fraction of infected, $d_i$ for each ABO blood group are reported in Table~\ref{tab1}, together with the frequencies of the groups in a control sample of 1881 hospitalized individuals, whose blood groups were collected in the same period of time of the 186 infected cases. 
The reduced size of the sample assures statistical confidence only for A and O groups (p-value $<0.05$), with AB being the less trust-wort with a p-value of 0.364 (see~\cite{32496734}).

Another work by Zietz~\textit{et al.}~\cite{Zietz2020}, collected information of  ABO groups for 682 infected individuals tested in New York-Presbyterian/Columbia University Irving Medical Center (NYP/CUIMC).

Finally, additional data was collected by two works that considered critically ill patients in Italy and Spain~\cite{ellinghaus2020genomewide} and in the United States~\cite{leaf2020abo}. Comparing blood frequencies of infected and control populations, a sub-representation of blood group O and an over-representation of group A was registered in both Italian and Spanish datasets, in accordance with the observations in Wuhan/Shenzhen and Denmark.
Data collected by Leaf et al.~\cite{leaf2020abo} are stratified by both ABO blood groups and ethnicity.

\subsubsection*{Data Analysis}

Once we have collected the availaible data on the number of infected people stratified by blood type, $d_i$, in a geographical area where the blood type distribution, $f_i$, is known we perform the comparison of these data with the model. According to Eqs. (\ref{yi}) and (\ref{pi}) we expect that 
\begin{equation}
d_i = y_o f_i\; e^{(p^{(k)}_i -\rho) \tau^*} \;\;\;\;\;\;\;\;\;\;\;\;\; i=1..k
\end{equation}
with $p^{(k)}_i = ({\bf W} \cdot \bar f)_i$ and $\tau^*$ the specific (reduced) time when the data have been collected. Therefore
\begin{equation}
\label{eq:pi}
\ln(d_i / f_i ) = K + p^{(k)}_i \tau^* \;\;\;\;\;\;\;\;\;\;\;\;\; i=1..k
\end{equation}
being $K$ a constant ($K = \ln{y_o} - \rho \tau^*$) not depending on $i$. We thus expect a linear relation between  $\ln(d_i / f_i ) $ and $p^{(k)}_i $. The error bars on $\ln(d_i / f_i ) $ are one standard deviation calculated by assuming a Poissonian distribution for the number of cases, thus $\Delta(ln(d_i / f_i ) )=1/\sqrt{n_i}$ being $n_i$ the number of infected persons with blood type $i$.

\subsection*{Cases where infection data are not stratified by blood type} 

To our knowledge, all the studies reporting the blood type stratification of the infection data are reported in Table~\ref{tab1}. To further test the blood type constrained infection rules proposed in \cite{breiman2020harnessing}, we have to abandon $p^{(k)}_i$ (the susceptibility per group) and focus on $p^{k}_T $ (the total susceptibility), for which a large amount of data is available.

\subsubsection*{The countries case}

The incidence of the different blood types in different nations can be found in Wikipedia \cite{ wiki:blood_type}, where a collection of data and the original sources references are reported. These frequencies $f$ are also listed in Table~\ref{tab2}, together with the country ISO code.
From these data we can calculate $p^{(2)}_{T}$ (keeping into account the Rh$\pm$ type) and $p^{(4)}_{T}$ (keeping into account only the ABO type)  according to Eq. (\ref{pi}) for each country. 
To estimate the uncertainty on $p^{(k)}_{T}$, we observe that the $f$ are reported with three significant digits. We therefore associate to each $f$ (which value is of the order of one) an error equal to $\Delta f = 10^{-3}$, and we estimate the upper limit of the error on $p^{(k)}_{T}$, which is bi-linear in $f$, as $\Delta  p^{(k)}_{T} \approx 2 \sqrt k \Delta f$.

The data of the contagion by country is taken from World Health Organization (WHO) Coronavirus Disease (COVID-19) Dashboard on date 12th of June 2020 \cite{WHO_data}.
To ensure statistical reliability, we selected only countries that had registered at least 2000 positive cases from the start of the epidemic. Requiring also to know the frequencies of both ABO and $Rh\pm$, we ended up with 78 countries, whose information is reported in Table~\ref{tab2}.

\begin{table*}
\centering
\caption{Percentages of blood groups ($f_i$) as reported in \cite{wiki:blood_type}, susceptibility, $p^{(k)}_T$ as given by  Eq. (\ref{pi}), and inverse characteristic time, $m$, of the exponential phase of  the infection for the analysed countries as derived for the fit to the observed data.}
\label{tab2}
\resizebox{0.75\textwidth}{!}{  
\begin{tabular}{|lcc|cccccccc|cc|ccc|}
\hline {\bf Country} & {\bf Cluster} & {\bf Code} &     $O_+$ &     $A_+$ &     $B_+$ &    $AB_+$ &    $O_-$ &    $A_-$ &    $B_-$ &   $AB_-$  & $p^{(2)}_{T}$  & $p^{(4)}_{T}$   &   $ m$ & $\Delta m$ & $R^2 $ \\
Argentina &        SA &      AR &  45.40 &  34.26 &   8.59 &   2.64 &  8.40 &  0.44 &  0.21 &  0.06 &  0.917 &  0.679 &    0.196 &  0.011 &  0.98 \\
Armenia &        AS &      AM &  29.00 &  46.30 &  12.00 &   5.60 &  2.00 &  3.70 &  1.00 &  0.40 &  0.934 &  0.618 &    0.143 &  0.025 &  0.97 \\
Australia &        AU &      AU &  40.00 &  31.00 &   8.00 &   2.00 &  9.00 &  7.00 &  2.00 &  1.00 &  0.846 &  0.660 &   0.223 &  0.005 &  0.99 \\
Austria &        EU &      AT &  30.00 &  37.00 &  12.00 &   5.00 &  6.00 &  7.00 &  2.00 &  1.00 &  0.866 &  0.612 &    0.216 &  0.033 &  0.98 \\
Bahrain &        AS &      BH &  48.48 &  19.35 &  22.61 &   3.67 &  3.27 &  1.33 &  1.04 &  0.25 &  0.945 &  0.635 &    0.356 &  0.054 &  0.97 \\
Bangladesh &        AS &      BD &  29.45 &  26.01 &  33.66 &   8.29 &  0.95 &  0.67 &  0.70 &  0.27 &  0.975 &  0.553 &   0.245 &  0.008 &  1.00 \\
Belgium &        EU &      BE &  38.00 &  34.00 &   8.60 &   4.10 &  7.00 &  6.00 &  1.50 &  0.80 &  0.870 &  0.647  &  0.209 &  0.019 &  0.99 \\
Bolivia &        SA &      BO &  51.53 &  29.45 &  10.11 &   1.15 &  4.39 &  2.73 &  0.54 &  0.10 &  0.928 &  0.680 &   0.148 &  0.012 &  0.95 \\
Bosnia and Herzegovina &        EU &      BA &  31.00 &  36.00 &  12.00 &   6.00 &  5.00 &  7.00 &  2.00 &  1.00 &  0.872 &  0.609 &    0.152 &  0.010 &  0.99 \\
Brazil &        SA &      BR &  36.00 &  34.00 &   8.00 &   2.50 &  9.00 &  8.00 &  2.00 &  0.50 &  0.843 &  0.653 &  0.272 &  0.015 &  0.98 \\
Bulgaria &        EU &      BG &  28.00 &  37.00 &  13.00 &   7.00 &  5.00 &  7.00 &  2.00 &  1.00 &  0.872 &  0.600 &   0.063 &  0.038 &  0.97 \\
Cameroon &        AF &      CM &  42.80 &  38.80 &  12.00 &   3.30 &  1.40 &  1.20 &  0.40 &  0.10 &  0.970 &  0.636 &   0.400 &  0.059 &  0.95 \\
Canada &        NA &      CA &  39.00 &  36.00 &   7.60 &   2.50 &  7.00 &  6.00 &  1.40 &  0.50 &  0.873 &  0.661 &    0.250 &  0.016 &  0.96 \\
Chile &        SA &      CL &  85.50 &   8.70 &   3.35 &   1.00 &  1.20 &  0.10 &  0.05 &  0.10 &  0.986 &  0.877 &     0.411 &  0.034 &  0.97 \\
China &        AS &      CN &  47.70 &  27.80 &  18.90 &   5.00 &  0.28 &  0.19 &  0.10 &  0.03 &  0.994 &  0.620 &    0.419 &  0.069 &  1.00 \\
Colombia &        SA &      CO &  61.30 &  26.11 &   2.28 &   1.47 &  5.13 &  2.70 &  0.70 &  0.31 &  0.919 &  0.754 &   0.218 &  0.015 &  0.97 \\
Croatia &        EU &      HR &  29.00 &  36.00 &  15.00 &   5.00 &  5.00 &  6.00 &  3.00 &  1.00 &  0.872 &  0.588 &    0.159 &  0.017 &  0.99 \\
Cuba &        SA &      CU &  45.80 &  33.50 &  10.20 &   2.90 &  3.60 &  2.80 &  1.00 &  0.20 &  0.930 &  0.654 &    0.220 &  0.020 &  0.95 \\
Czechia &        EU &      CZ &  27.00 &  36.00 &  15.00 &   7.00 &  5.00 &  6.00 &  3.00 &  1.00 &  0.872 &  0.583 &  0.287 &  0.029 &  0.98 \\
Congo &        AF &      CD &  59.50 &  21.30 &  15.20 &   2.40 &  1.00 &  0.30 &  0.20 &  0.10 &  0.984 &  0.685 &   0.123 &  0.010 &  0.99 \\
Denmark &        EU &      DK &  35.00 &  37.00 &   8.00 &   4.00 &  6.00 &  7.00 &  2.00 &  1.00 &  0.866 &  0.643 &   0.320 &  0.049 &  0.99 \\
Dominican Republic &        SA &      DO &  46.20 &  26.40 &  16.90 &   3.10 &  3.70 &  2.10 &  1.40 &  0.20 &  0.931 &  0.630  &  0.295 &  0.047 &  0.94 \\
Ecuador &        SA &      EC &  75.00 &  14.00 &   7.10 &   0.50 &  2.38 &  0.70 &  0.30 &  0.02 &  0.967 &  0.802 &    0.471 &  0.028 &  0.99 \\
Egypt &        AF &      EG &  52.00 &  24.00 &  12.40 &   3.80 &  5.00 &  2.00 &  0.60 &  0.20 &  0.928 &  0.672 &   0.158 &  0.029 &  0.97 \\
El Salvador &        SA &      SV &  62.00 &  23.00 &  11.00 &   1.00 &  1.00 &  1.00 &  0.70 &  0.30 &  0.971 &  0.706 &    0.148 &  0.007 &  0.98 \\
Ethiopia &        AF &      ET &  39.00 &  28.00 &  21.00 &   5.00 &  3.00 &  2.00 &  1.00 &  1.00 &  0.935 &  0.593  &  0.087 &  0.002 &  1.00 \\
Finland &        EU &      FI &  28.00 &  35.00 &  16.00 &   7.00 &  5.00 &  6.00 &  2.00 &  1.00 &  0.880 &  0.584 &  0.127 &  0.007 &  0.98 \\
France &        EU &      FR &  36.00 &  37.00 &   9.00 &   3.00 &  6.00 &  7.00 &  1.00 &  1.00 &  0.872 &  0.647 &   0.251 &  0.007 &  0.99 \\
Germany &        EU &      DE &  35.00 &  37.00 &   9.00 &   4.00 &  6.00 &  6.00 &  2.00 &  1.00 &  0.872 &  0.637 &  0.230 &  0.022 &  0.99 \\
Ghana &        AF &      GH &  53.80 &  17.60 &  18.30 &   2.80 &  4.50 &  1.30 &  1.30 &  0.20 &  0.928 &  0.668 &   0.091 &  0.013 &  0.98 \\
Greece &        EU &      GR &  37.40 &  32.90 &  11.00 &   3.70 &  7.00 &  5.00 &  2.00 &  1.00 &  0.872 &  0.631 &   0.282 &  0.063 &  0.94 \\
Guinea &        AF &      GN &  46.88 &  21.64 &  22.86 &   4.52 &  2.00 &  0.90 &  1.00 &  0.20 &  0.961 &  0.621 &   0.094 &  0.005 &  0.99 \\
Honduras &        SA &      HN &  57.50 &  27.00 &   7.80 &   2.50 &  2.70 &  1.70 &  0.60 &  0.20 &  0.951 &  0.702 &  0.128 &  0.015 &  0.98 \\
Hungary &        EU &      HU &  27.00 &  33.00 &  16.00 &   8.00 &  5.00 &  7.00 &  3.00 &  1.00 &  0.866 &  0.577 &   0.132 &  0.006 &  1.00 \\
India &        AS &      IN &  27.85 &  20.80 &  38.14 &   8.93 &  1.43 &  0.57 &  1.79 &  0.49 &  0.959 &  0.565 &    0.191 &  0.005 &  1.00 \\
Indonesia &        AS &      ID &  36.82 &  25.87 &  28.85 &   7.96 &  0.18 &  0.13 &  0.15 &  0.04 &  0.995 &  0.572 &  0.292 &  0.044 &  0.97 \\
Iran &        AS &      IR &  33.50 &  27.00 &  22.20 &   7.00 &  4.00 &  3.00 &  2.50 &  0.80 &  0.908 &  0.575 &    0.054 &  0.004 &  1.00 \\
Iraq &        AS &      IQ &  32.10 &  25.00 &  25.60 &   7.40 &  3.60 &  2.70 &  2.70 &  0.90 &  0.911 &  0.567 &    0.092 &  0.003 &  1.00 \\
Ireland &        EU &      IE &  47.00 &  26.00 &   9.00 &   2.00 &  8.00 &  5.00 &  2.00 &  1.00 &  0.866 &  0.672 &   0.307 &  0.018 &  1.00 \\
Israel &        AS &      IL &  32.00 &  34.00 &  17.00 &   7.00 &  3.00 &  4.00 &  2.00 &  1.00 &  0.910 &  0.582 &    0.180 &  0.012 &  1.00 \\
Italy &        EU &      IT &  39.00 &  36.00 &   7.50 &   2.50 &  7.00 &  6.00 &  1.50 &  0.50 &  0.872 &  0.661 &    0.389 &  0.020 &  0.99 \\
Ivory Coast &        AF &      CI &  46.50 &  22.50 &  22.50 &   4.30 &  2.00 &  1.00 &  1.00 &  0.20 &  0.960 &  0.619 &    0.073 &  0.016 &  0.97 \\
Japan &        AS &      JP &  29.90 &  39.80 &  19.90 &   9.90 &  0.15 &  0.20 &  0.10 &  0.05 &  0.995 &  0.570  &  0.150 &  0.023 &  0.99 \\
Kazakhstan &        AS &      KZ &  30.70 &  29.80 &  24.20 &   8.30 &  2.30 &  2.20 &  1.80 &  0.70 &  0.935 &  0.560  &  0.085 &  0.014 &  0.98 \\
Kenya &        AF &      KE &  45.60 &  25.20 &  21.28 &   4.20 &  1.80 &  1.00 &  0.90 &  0.02 &  0.964 &  0.614  &  0.277 &  0.045 &  0.98 \\
Luxembourg &        EU &      LU &  35.00 &  37.00 &   9.00 &   4.00 &  6.00 &  6.00 &  2.00 &  1.00 &  0.872 &  0.637 &  0.207 &  0.022 &  0.97 \\
Malaysia &        AS &      MY &  34.32 &  30.35 &  27.37 &   7.46 &  0.17 &  0.15 &  0.14 &  0.04 &  0.995 &  0.563 &   0.220 &  0.018 &  0.96 \\
Mexico &        SA &      MX &  59.09 &  26.23 &   8.53 &   1.73 &  2.73 &  1.21 &  0.40 &  0.08 &  0.958 &  0.708  &  0.225 &  0.031 &  0.97 \\
Moldova &        EU &      MD &  28.50 &  31.80 &  17.60 &   7.00 &  5.00 &  6.00 &  3.00 &  1.10 &  0.872 &  0.574 &  0.168 &  0.008 &  0.99 \\
Morocco &        AF &      MA &  42.30 &  29.80 &  14.30 &   4.10 &  4.50 &  3.10 &  1.50 &  0.40 &  0.914 &  0.625 &   0.176 &  0.009 &  1.00 \\
Nepal &        AS &      NP &  35.20 &  28.30 &  27.10 &   8.60 &  0.30 &  0.20 &  0.20 &  0.10 &  0.992 &  0.567  &  0.135 &  0.025 &  0.96 \\
Netherlands &        EU &      NL &  39.50 &  35.00 &   6.70 &   2.50 &  7.50 &  7.00 &  1.30 &  0.50 &  0.864 &  0.669  &  0.439 &  0.036 &  0.99 \\
Nigeria &        AF &      NG &  51.30 &  22.40 &  20.70 &   2.60 &  1.60 &  0.70 &  0.60 &  0.10 &  0.971 &  0.640 &   0.155 &  0.020 &  0.92 \\
Macedonia &        EU &      MK &  30.00 &  34.00 &  15.00 &   6.00 &  5.00 &  6.00 &  3.00 &  1.00 &  0.872 &  0.588 &    0.224 &  0.034 &  0.97 \\
Norway &        EU &      NO &  33.20 &  41.60 &   6.80 &   3.40 &  5.80 &  7.40 &  1.20 &  0.60 &  0.873 &  0.661 &    0.304 &  0.047 &  0.96 \\
 Pakistan &        AS &      PK &  26.63 &  21.60 &  34.40 &   9.52 &  2.17 &  1.66 &  3.57 &  0.45 &  0.928 &  0.557  &  0.075 &  0.005 &  0.99 \\
Peru &        SA &      PE &  70.00 &  18.40 &   7.80 &   1.60 &  1.40 &  0.50 &  0.28 &  0.02 &  0.978 &  0.761 &    0.385 &  0.021 &  0.99 \\
Philippines &        AS &      PH &  45.90 &  22.90 &  24.90 &   5.97 &  0.10 &  0.10 &  0.10 &  0.03 &  0.997 &  0.608  &  0.353 &  0.027 &  0.95 \\
Poland &        EU &      PL &  31.00 &  32.00 &  15.00 &   7.00 &  6.00 &  6.00 &  2.00 &  1.00 &  0.872 &  0.594 &   0.233 &  0.022 &  0.99 \\
Portugal &        EU &      PT &  36.20 &  39.80 &   6.60 &   2.90 &  6.10 &  6.80 &  1.10 &  0.50 &  0.876 &  0.666  &  0.389 &  0.035 &  0.99 \\
Romania &        EU &      RO &  28.00 &  36.50 &  13.60 &   6.80 &  5.00 &  6.50 &  2.40 &  1.20 &  0.872 &  0.594 &   0.184 &  0.009 &  0.99 \\
Russian Federation &        EU &      RU &  28.00 &  30.00 &  20.00 &   7.00 &  4.90 &  5.80 &  3.20 &  1.10 &  0.872 &  0.565 &   0.185 &  0.004 &  1.00 \\
Saudi Arabia &        AS &      SA &  47.80 &  23.90 &  17.00 &   4.00 &  4.00 &  2.00 &  1.00 &  0.30 &  0.932 &  0.638 &   0.190 &  0.037 &  0.96 \\
Serbia &        EU &      RS &  31.92 &  35.28 &  12.60 &   4.20 &  6.08 &  6.72 &  2.40 &  0.80 &  0.866 &  0.610 &    0.143 &  0.004 &  1.00 \\
Singapore &        AS &      SG &  44.70 &  23.90 &  24.50 &   5.60 &  0.60 &  0.30 &  0.30 &  0.10 &  0.987 &  0.604 &    0.178 &  0.066 &  0.98 \\
South Africa &        AF &      ZA &  39.00 &  32.00 &  12.00 &   3.00 &  6.00 &  5.00 &  2.00 &  1.00 &  0.880 &  0.629 &   0.208 &  0.024 &  0.98 \\
South Korea &        AS &      KR &  27.90 &  33.87 &  26.92 &  10.98 &  0.10 &  0.13 &  0.08 &  0.02 &  0.997 &  0.548  &  0.195 &  0.011 &  0.97 \\
Spain &        EU &      ES &  35.00 &  36.00 &   8.00 &   2.50 &  9.00 &  7.00 &  2.00 &  0.50 &  0.849 &  0.652 &    0.377 &  0.011 &  0.99 \\
Sudan &        AF &      SD &  48.00 &  27.70 &  15.20 &   2.80 &  3.50 &  1.80 &  0.80 &  0.20 &  0.941 &  0.642  &  0.110 &  0.003 &  1.00 \\
Sweden &        EU &      SE &  32.00 &  37.00 &  10.00 &   5.00 &  6.00 &  7.00 &  2.00 &  1.00 &  0.866 &  0.625  &  0.303 &  0.010 &  0.99 \\
Switzerland &        EU &      CH &  35.00 &  38.00 &   8.00 &   4.00 &  6.00 &  7.00 &  1.00 &  1.00 &  0.872 &  0.650 &    0.241 &  0.044 &  0.98 \\
Thailand &        AS &      TH &  40.80 &  16.90 &  36.80 &   4.97 &  0.20 &  0.10 &  0.20 &  0.03 &  0.995 &  0.605 &    0.270 &  0.006 &  0.99 \\
Turkey &        AS &      TR &  29.80 &  37.80 &  14.20 &   7.20 &  3.90 &  4.70 &  1.60 &  0.80 &  0.902 &  0.596 &    0.201 &  0.036 &  0.98 \\
Ukraine &        EU &      UA &  32.00 &  34.00 &  15.00 &   5.00 &  5.00 &  6.00 &  2.00 &  1.00 &  0.880 &  0.597 &   0.229 &  0.016 &  0.99 \\
United Arab Emirates &        AS &      AE &  44.10 &  21.90 &  20.90 &   4.30 &  4.30 &  2.10 &  2.00 &  0.40 &  0.920 &  0.618 &    0.135 &  0.007 &  0.98 \\
United Kingdom &        EU &      BG &  38.00 &  32.00 &   8.00 &   3.00 &  9.00 &  7.00 &  2.00 &  1.00 &  0.846 &  0.653 &    0.230 &  0.008 &  0.99 \\
United States &        NA &      US &  37.40 &  35.70 &   8.50 &   3.40 &  6.60 &  6.30 &  1.50 &  0.60 &  0.872 &  0.649 &   0.306 &  0.016 &  0.97\\
Venezuela &        SA &      VE &  58.30 &  28.20 &   5.60 &   1.90 &  3.70 &  1.80 &  0.40 &  0.10 &  0.944 &  0.721 &  0.171 &  0.029 &  0.84 \\
\hline
\end{tabular}
}
\end{table*}

In particular, the WHO data reports the number of new infections per day, $\dot D(t)$. From this quantity, we easily obtain the cumulative number of people that have been infected as a function of time, $D(t) = \int_o^t \dot D(t') dt'$. The cumulative, rather than directly with $\dot D(t)$ allows us to work on cleaner and more solid data because the day-by-day fluctuations are averaged out in the long run. Coming back to the model, at short times the cumulative $D_T(\tau)$ is directly related to $y_T(\tau)$, which short-time expansion is reported in Eq. (\ref{yyyy}). Overall, the COVID19 infection is characterized by low mortality but a high infectivity rate together with both long incubation and recovery periods. These conditions assure that limiting our analysis to the early stages of the infection, the cumulative can be regarded as a good proxy for the number of infected.

\begin{figure*}[t]
\centering
\includegraphics[width=\linewidth]{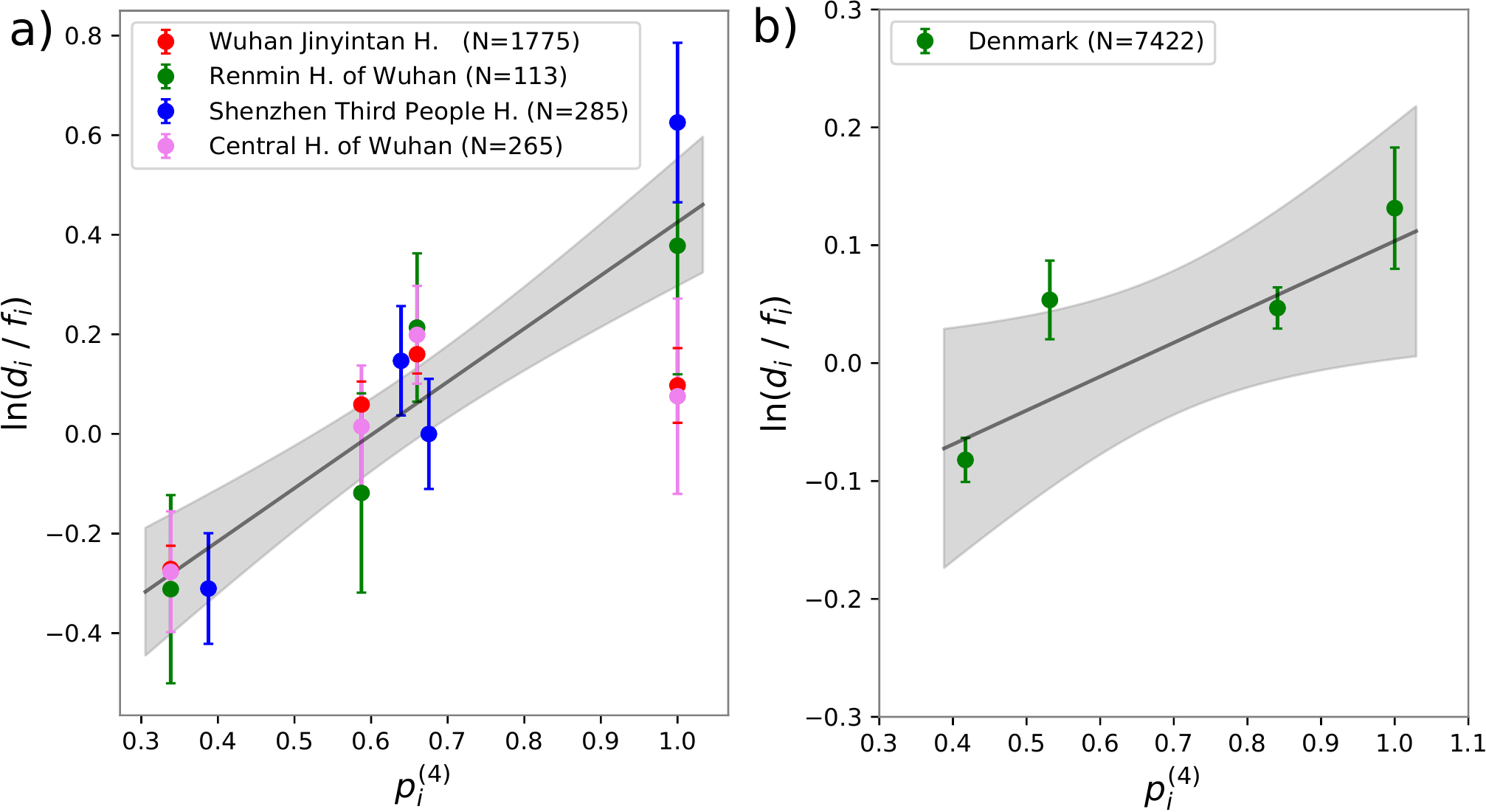}
\caption{ \textbf{a)} Ratio between frequencies of infected people for each ABO group, $d_i$ over frequency of the group on the  whole population, $f_i$ as measured in three hospitals of the Wuhan/Shenzhen region~\cite{zhao2020relationship,Li2020} versus its theoretical prediction, $p_i^{(4)}$  obtained from Eq.~(\ref{eq:pi}) using the ABO set of rules and the blood type frequencies in Table~\ref{tab1}. Black line represents the best solution of a linear fit.
\textbf{b)}  Ratio between frequencies of infected people for each ABO groups, $d_i$ over frequency of the group on the  whole population, $f_i$ as measured in Denmark~\cite{Barnkob2020} versus its theoretical prediction, $p_i^{(4)}$  obtained from Eq.~(\ref{eq:pi}) using the ABO set of rules and the blood type frequencies in Table~\ref{tab1}. Black line represents the best solution of a linear fit.
} 
\label{fig_cinesi}
\end{figure*}

For each country reported in Table~\ref{tab2},  we manually selected the time interval corresponding to the initial stage of the infection. In that range, we performed an unsupervised fit using the function:
\begin{equation}
Y(t) = \begin{cases}
m_0 t & \text{if} \qquad t<t_0\\
m_0 t_0 + A\left( e^{m(t-t_0)} -1 \right) & \text{if} \qquad t>t_0\\
\end{cases}
\end{equation}
with $t_o$, $m_o$, $m$ and $A$ as free parameters. In this expression, $A$ is a scaling factor and $m$ represents the inverse of the characteristic time of the infection exponential growth, which is the quantity that we are looking at. The linear term, observed at the very beginning of the infection curve of different (not all) countries, can be rationalized by assuming that at the very early stage of the infection, people arrive from other countries, spreading the virus before the real exponential growth appears. The constant $m_0$ is the ``arrival rate" of infected people (which is reasonable to assume constant in time), and we do not expect any correlation between $m$ and $m_0$,
because $m$ is an intrinsic characteristic of the population,
and $m_0$ only depend on the arrival from abroad. Besides the specific meaning of the different parameters, 
our aim is to derive the value of $m$, to compare this value with the prediction of the model $\pi_T^{(k)}$.  

In Table~\ref{tab2}, we report the values of $m$ with its statistical error $\Delta m$ as derived from the fit.
Examples of the fits are reported in the Supporting Information.
In this section, we compare the outcome of the model with the available observational data (presented in the previous section). Here \textit{model} indicates the combination of the generalized SIR  \textit{and} the infection rules hypothesized in  \cite{breiman2020harnessing}.

\section*{Results}

\subsection*{Cases where infection data are stratified by blood type} 

At first, we considered the Wuhan/Shenzhen data.
In the papers by Zhao et al. \cite{zhao2020relationship} and Li et al.~\cite{Li2020} data of the COVID-19 contagion are reported, stratified by ABO blood type (no info on the Rh$\pm$ type are given). In particular, for each of the four investigated hospitals, the authors provide the frequencies of the ABO blood types of the local population, $\bar f = (f_0,f_A,f_B,f_{AB})$ (three of them are the same as the hospitals are in the same city). The authors also report the fraction of cases for each blood ABO type: $\bar d = (d_0,d_A,d_B,d_{AB})$. These values are summarised in table \ref{tab1}.

As discussed before, the model predicts a linear relation between  $\ln(d_i / f_i ) $ and $p^{(4)}_i $. In Fig. (\ref{fig_cinesi}a) we report a plot of these two quantities. The different colors represent the different datasets.  The black line is the result of the least square fitting to all the data. Finally, the light grey indicates the $\pm$ one standard deviation area. 

The Pearson correlation coefficient for the N=16 data points is 0.82. The corresponding p-value ($10^{-4}$) gives us confidence that the model is compatible with the observations.

Further data, with higher statistics, was collected by~\cite{Barnkob2020}, based on Danish individuals found positive to COVID-19 (see Table~\ref{tab1}).  Figure~\ref{fig_cinesi}b shows once again the logarithm of the ratio between frequencies of infected people for each ABO group, $d_i$ over frequency of the group on the healthy population, $f_i$ against the expected susceptibility, $p_i^{(4)}$. Even in this case, a correlated trend is appreciable. 
The Pearson correlation coefficient is 0.85 when considering all four blood groups (N=4, p-value 0.15).
If we do not consider B group data, the only one that is not statistically significant (see ~\cite{Barnkob2020}), the correlation increases to 0.99 (N=3, p-value 0.08).
Combining the Wuhan/Shenzen datasets with the Denmark one, we obtained an overall correlation of 0.78 (N=20, p-value $<10^{-4}$).

To our knowledge, the five datasets, considered so far, are the only available ones for which (i) not only severely ill patients were considered and (ii) the control population was not composed of hospital patients. This allows us to test the contribution of the blood type to the sole infection transmission. 

Next, we took into consideration all datasets in Table~\ref{tab1}.

\begin{figure*}[t]
\centering
\includegraphics[width=\linewidth]{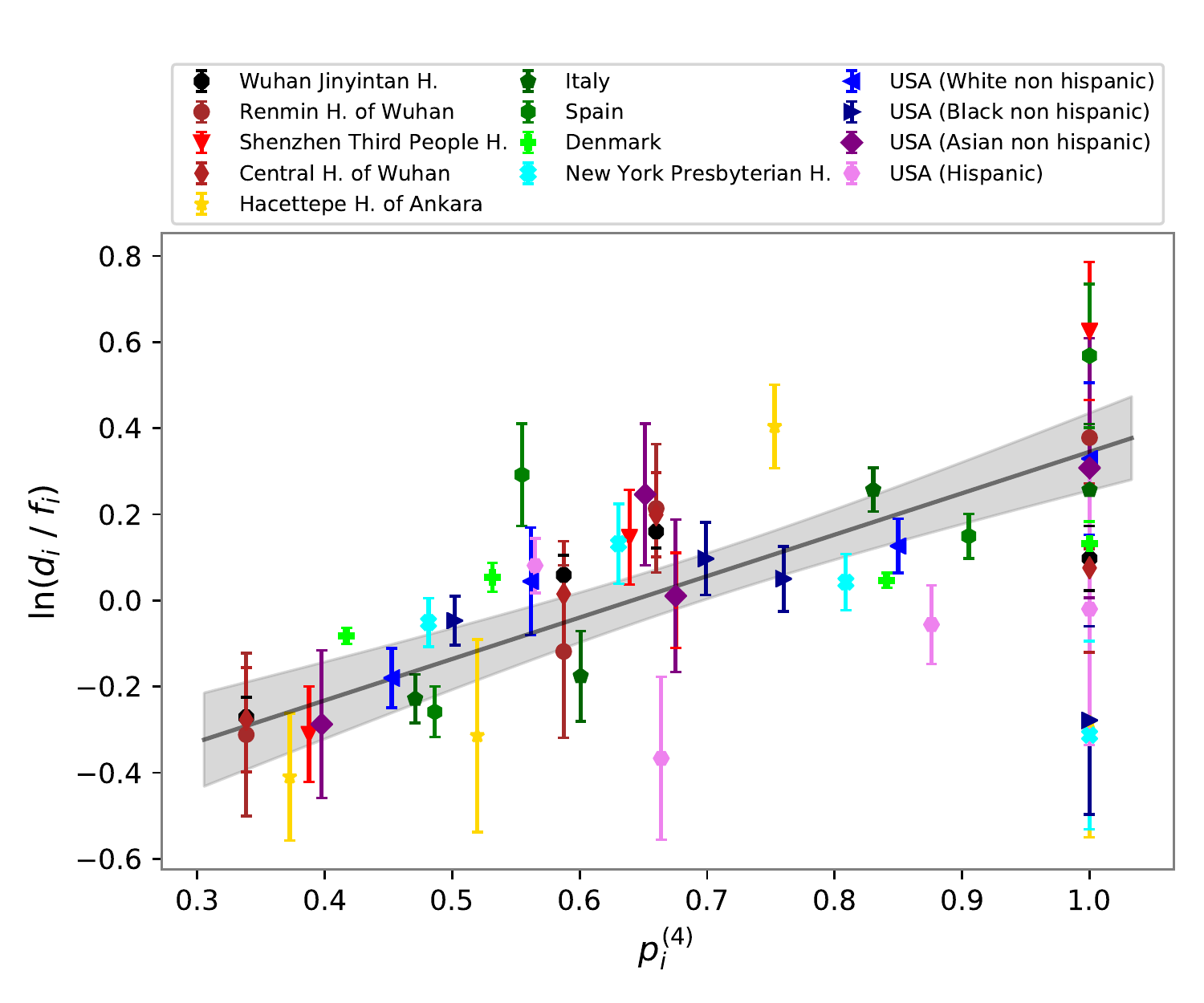}
\caption{ Ratio between frequencies of infected people for each ABO, $d_i$ over frequency of the group on the  whole population, $f_i$  versus its theoretical prediction, $p_i^{(4)}$  obtained from Eq.~(\ref{eq:pi}) using the ABO set of rules and all the blood type frequencies in Table~\ref{tab1}. Black line represents the best solution of a linear fit.} 
\label{fig_all}
\end{figure*}

In Figure~\ref{fig_all}, the logarithm of the ratio between the number of infected for each group over the group frequency of the local population ($\ln (d_i/f_i)$)  is compared with the 4-group susceptibility, $p_i^{(4)}$ computed according to Eq.~\ref{pi_i}  and using the data in Table~\ref{tab1}.
We observe a linear correlation of 0.53 with a p-value smaller than $10^{-4}$.

\subsection*{Cases where infection data are not stratified by blood type}

\subsubsection*{Europe} 

The quantities $\pi^{(k)}_T $ represent the inverse of the characteristic time of the infection (of the whole population) at its initial exponential stage. Therefore, they can be directly compared with the quantities $m$ previously discussed. Since the slopes $m$ are measured plotting the data as a function of the real-time $t$, while $\pi^{(k)}_T $ are proportional to the scaled time $\tau = \beta t$, we expect the relation $m = \beta \pi^{(k)}_T=\beta p^{(k)}_T -\gamma$. Therefore, from a plot of $m$ vs. $p^{(k)}_T $, we can both check the validity of the model and extract the parameters $\beta$ and $\gamma$. 

\begin{figure*}[t]
\centering
\includegraphics[width=\linewidth]{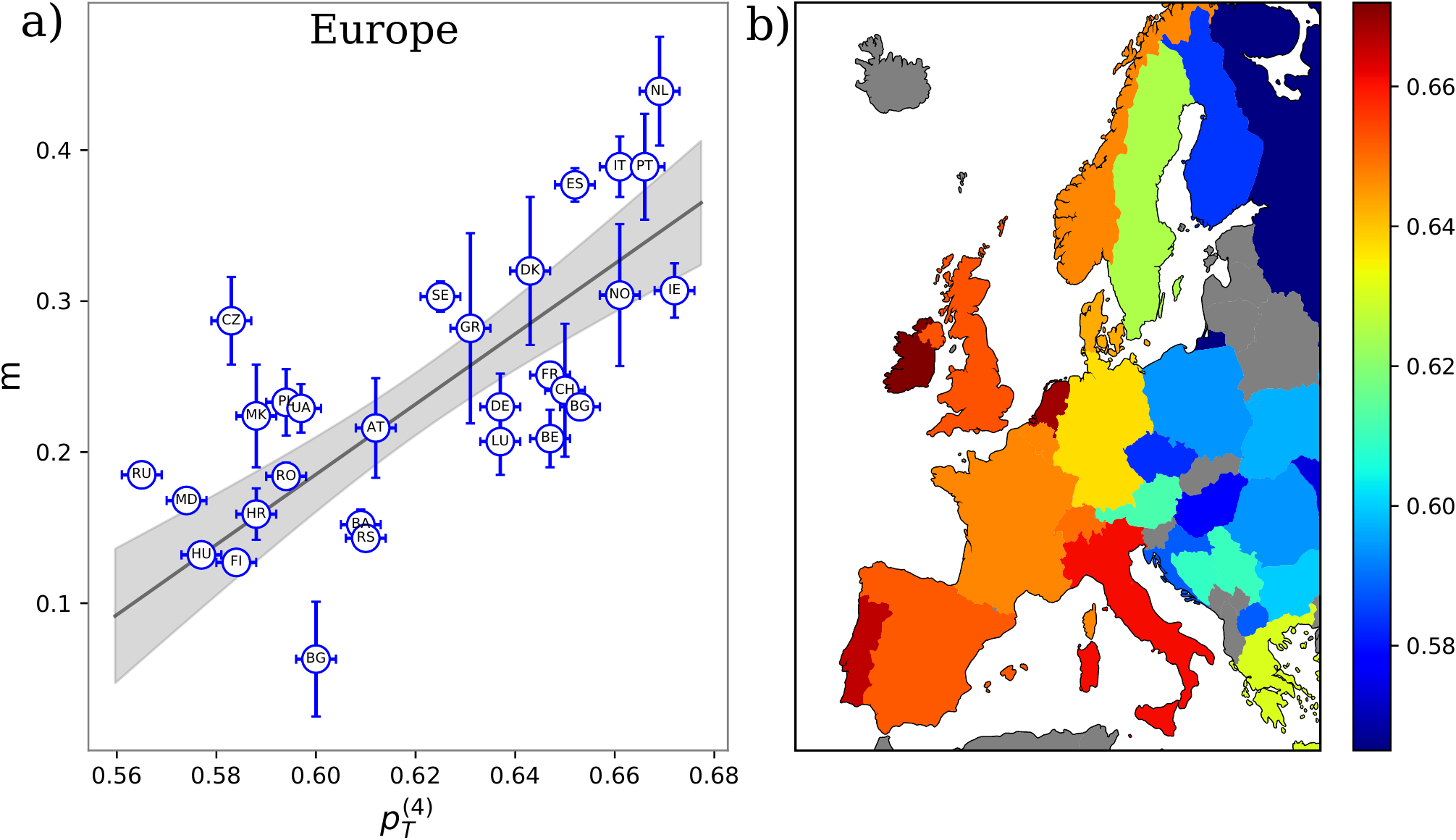}
\caption{\textbf{a)} Inverse characteristic time of the epidemic exponential phase extracted from cumulative infection curves, $m$, vs theoretical prediction, $p_T^{(4)}$  obtained from Eq.~(\ref{eq:p_k}) using the ABO set of rules and the blood type frequencies in Table~\ref{tab2}. Each dot corresponds to one of the 29 analyzed countries in the European region named according to the 2-letter ISO code and reported in Table~\ref{tab2}. The black line represents the best solution of a linear fit performed with the York method and the grey shaded area is the $\pm$ one standard deviation confidence band.
\textbf{b)}Map representation of European countries. Each country is colored according to its $p_T^{(4)}$ susceptibility value obtained from Eq.~(\ref{eq:p_k}) using the ABO set of rules and the blood type frequencies reported in Table~\ref{tab2}. $p_T^{(4)}$ values increase going from blue to red. Gray countries have not been considered due to a lack of either blood or infection information. The map shows large variability in the susceptibility $p_T^{(4)}$ (that ranges from 0.56 to 0.68) and a clear east-to-west gradient. The increase of susceptibility going west is a direct consequence of the tendency of increase of the 0 blood type in this direction: the more one blood type dominates, the higher is the susceptibility.
}
\label{fig_europa}
\end{figure*}

At first, we compared the $m$ found for the 29 countries of the European area with the corresponding $p^{(k)}_T $s computed starting from the frequencies of the different blood types found in each country and reported in Table~\ref{tab2}.      
While a contagion scheme based only on the Rhesus group rules ($p^{(2)}_T$) does not explain the observed trends of the epidemics, a high Pearson correlation (0.71) is present between data and model predictions when contagions are driven by the ABO group rules ($p^{(4)}_T$). 
Indeed, as shown in Figure~\ref{fig_europa}a, countries with higher susceptibility also present, on average, a higher $m$ value. A linear fit of the data allows us to extract the overall values of infection ($\beta$) and recovery ($\gamma$) rates that if combined yield a value of  $R_0$ of $\sim 2$ in excellent agreement with the current estimates of reproduction number in the early stages of the outbreak~\cite{Kucharski2020}. 
We note that points tend to form two clusters. Interestingly, retrieving the geographic information, we see a clear east-to-west gradient of the susceptibility, which drives back to the different geographical distribution of the ABH phenotypes~\cite{Mourant1952}. Figure~\ref{fig_europa}b clearly shows at a glance this trend for  the four-group susceptibility, $p_T^{(4)}$. 


To test the obtained linear trends, we performed an F-test assuming a constant slope as null-hypothesis. 
In Table~\ref{tab3}, p-values for both the Pearson coefficient and F-statistic are reported. Note that the significance of $p^{(2)}_T$ is just an artifact of the data disposition, which tend to cluster around the values of $p_T^{(2)} = 0.87$, thus yielding a high slope. Reversing the axis and repeating the linear fit, one obtains a value of F of 1.28 which has a p-value above the threshold.    

\subsubsection*{Asia} 

\begin{figure}[t]
\centering
\includegraphics[width=\linewidth]{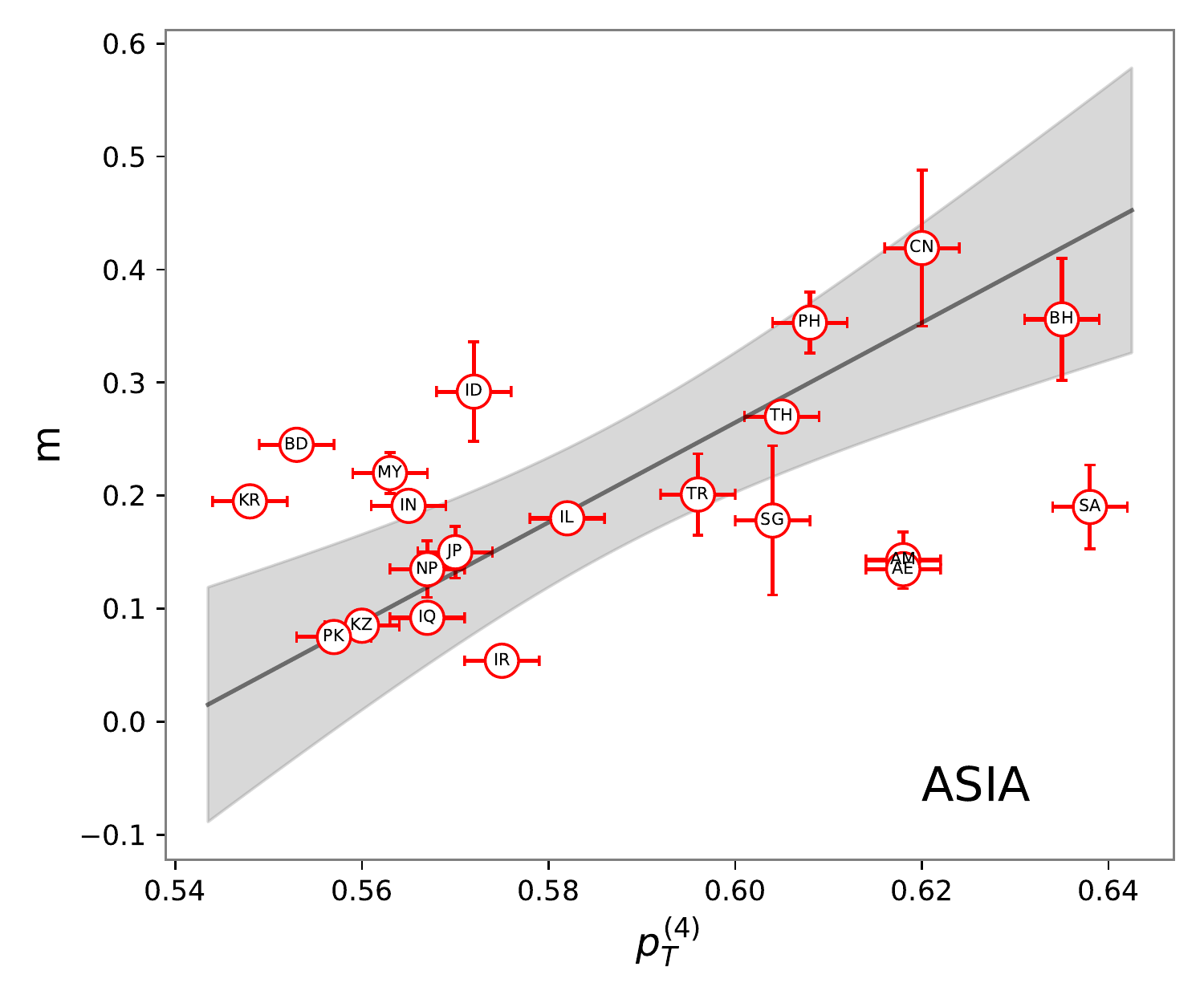}
\caption{Inverse characteristic time of the epidemic exponential phase extracted from cumulative infection curves, $m$ vs theoretical prediction, $p_T^{(4)}$  obtained from Eq.~(\ref{eq:p_k})  using the ABO sets of rules and using the blood type frequencies in Table~\ref{tab2}. Each dot corresponds to one of the 21 analyzed countries in the Asiatic region named according to the 2-letter ISO code and reported in Table~\ref{tab2}. The black line represents the best solution of a linear fit.}
\label{fig_asia}
\end{figure}

Repeating the same analyses carried out for European countries with those in the Asia region, we found an overall similar trend (see Figure~\ref{fig_asia}). In particular,  $p_T^{(4)}$ values present a non-random linear correlation with the inverse characteristic times, $m$, of 0.45. The corresponding p-values are below the threshold of 0.05 for the 21 Asiatic countries.

Testing the linear fits with the F-test, we can reject the null hypothesis for all three cases, with the usual threshold level of 0.05.
All results are again summarized in Table~\ref{tab3}.

\subsubsection*{South America and Africa} 
Finally, we consider two other distinct regions, i.e. South America and Africa, both characterized by a still exponentially proliferating infection.
The former exhibits a trend similar to the Europe one, characterized by an unsubstantial contribution of the Rhesus group rules, while a correlation of 0.62 is present between the $m$ coming from the fitting of the infection curves and $p_T^{(4)}$. The infection and recovering rates obtained by a linear fit are both smaller than those found in the Eurasian region. However, they combine to give a value of $R_0$ of $\sim 2$.
Both the p-value of the Pearson correlation and that an F-test on the linear fit are below the significance threshold of 0.05 for all set of rules except for $p_T^{(2)}$ one. 
Results for the African region instead show no meaningful correlation.  As it is the slope of the best linear fit which does not pass an F-test with zero slope as a null hypothesis. 

\begin{table*}[t]
\centering
\caption{Main results for the different countries aggregation.}
\label{tab3}
\begin{tabular}{|cc|cc||c||cc|cc|}
\hline {\bf Continent} &   & {$\beta$} & {$\gamma$}& {$\rho$}  & Pearson  & p-value  & F & p-value \\
\hline 
\hline 
Europe (EU)  & $p^{(2)}_T $ & -35.39 & -31.01 & -  &-0.26 & 0.17 & 22.4 &$<10^{-4}$ \\
 (29 countries)& $p^{(4)}_T $  & 2.31 & 1.20 & 0.52 & 0.71 &$<10^{-4}$  & 42.5 & $<10^{-4}$\\
 \hline \hline 
 Asia (AS) & $p^{(2)}_T $  & 2.06 & 1.80& - & 0.54 & 0.01& 19.5 & $<10^{-3}$\\
 (21 countries)& $p^{(4)}_T $  &  4.42 & 2.38 & 0.54 & 0.45 &0.04  & 1.3 &0.05\\
\hline \hline 
South America (SA)& $p^{(2)}_T $  & -0.46 & -0.62& -  & 0.29 & 0.34 &0.79 &  0.39\\
(13 countries) & $p^{(4)}_T $  & 1.24 & 0.67 & 0.54  & 0.62  & 0.02 & 8.1 &0.02 \\
 \hline \hline 
Africa (AF)& $p^{(2)}_T $  & 0.37 & 0.25 & 0.74  & -0.09 & 0.78  & 0.5 & 0.5 \\
 (12 countries)& $p^{(4)}_T $  & 0.52 & 0.22 & 0.42  & -0.07 & 0.82 & 1.2 &  0.3\\
\hline
\end{tabular}
\end{table*}

\subsubsection*{Final comments on the countries cases} 
Overall, we found a statistically significant correlation among the $m$ and $p^{(4)}_T$  data for Europe, Asia, and South Americas, taken individually. The reason for analyzing separately these continents lies in the fact that we expect that beside blood type distribution other factors affect the infectivity onset and initial growth rate. The lifestyle and the local climate are certainly some of them. We have therefore considered countries aggregations that preserve at the best these two aspects. To check this hypothesis, we have analyzed (see Table~\ref{tab4}) what is the effect of adding North America and/or Australia to the European countries. As can be seen in Table~\ref{tab4}, the Pearson correlation does not change significantly in these cases. A rather worst result is obtained by considering Asia and Europe together, although the Pearson correlation still maintains a high degree of statistical significance (p-value better than $10^{-4}$ ). The correlation becomes even worst when considering the whole world (Pearson 0.43) but also in this case there is a great advantage (p-value better than $10^{-3}$) with respect to the null hypothesis (no-correlation between $m$ and $p^{(k)}_T$). 
We conclude that the model presented here is compatible with the existing data.

As far as the effect of temperature, humidity, etc. it is tempting to speculate on the obtained results: \textit{i)} Europe and Asia share a good correlation, as well as the values of $\beta$ and $\gamma$; \textit{ii)} South America has again a good correlation, but its $\beta$ and $\gamma$ are smaller than in Europe and Asia; \textit{iii)} Africa shows a bad correlation. These three points could be rationalized remembering that, at the pandemic initial stage Europe and Asia were in their wintertime, South America in the summertime, while African countries experienced different climate situations.  

\begin{table}[t]
\centering
\caption{Pearson correlations between $m$ and $p^{(4)}_T$  for different continent aggregations.}
\label{tab4}
\begin{tabular}{|cc|cc|}
\hline {\bf Aggregation}& Countries &  Pearson  & p-value \\
\hline 
\hline 
Europe and North America & 31 &   0.70 & $<10^{-4}$  \\
\hline 
\hline 
Europe and Australia & 30& 0.69 & $<10^{-4}$  \\
\hline 
\hline 
Europe and  Asia & 50& 0.62 & $<10^{-4}$\\
\hline 
\hline 
Temperate (North)  & 48& 0.63 & $<10^{-4}$\\
\hline 
\hline 
Tropical  & 26 & 0.24 & 0.24\\
\hline 
\hline 
Temperate (Sud)  & 4& 0.97 & 0.03\\
\hline 
\hline 
World & 78 & 0.43 & $<10^{-3}$\\
\hline 
\end{tabular}

\end{table}

\section{Discussion}

Most of the proteins that decorate cell membranes are bound to glycans~\cite{10.1093/glycob/cww086}. The presence of those carbohydrate chains provides a further channel of interaction between proteins, besides the usual direct protein-protein one, and evolved to play a large array of life-sustaining functions including support, signaling, protein folding, and protection. 

It has been suggested that protection against pathogens was the driving force that favored the evolution of the complex landscape of glycan interactions (see e.g. \cite{10.1093/glycob/cwm005} for a more detailed discussion).
Since viruses do not have genes for glycan synthesis or modification, they inherit host cell glycans after each round of replication in a new host.
This means that the host cell in which the virus last replicated generated the glycans on viral glycoproteins~\cite{10.1093/glycob/cwm005}.

A mutation in the host population having as an outcome the loss of a glycan modification could then provide a selective advantage to the glycan-lacking subpopulation. In fact,  pathogens using that glycan as a receptor would not be able to invade the host cells anymore. Moreover, the host can develop specific antibodies against the abolished glycan~\cite{Patel2016}. 
Human blood groups constitute an important example.
Individuals having O blood group lack A or B antigens and when presented with glycan motifs similar to either A or B antigens, they develop anti-A and anti-B antibodies. Individuals with A or B blood group develop either anti-B or anti-A antibodies, respectively.  On the other hand, people with subgroup AB are not able to develop such antibodies. 
According to those `rules', in case of an infection,  one would expect that looking at the blood type of the people found infected by the virus, group O should be under-represented with respect to its occurrence in the whole population. 
This feature has been indeed observed by Zhao and coworkers~\cite{zhao2020relationship} for the COVID-19 outbreak, caused by the novel SARS-CoV-2 coronavirus.  Notably, a similar behavior was found in the hospital outbreak of SARS in Hong Kong in 2003~\cite{Cheng2005} and in that of the West Nile virus in Greece in 2010~\cite{Politis2016}. Moreover, both SARS-CoV and SARS-CoV-2 S protein trimers are covered by an extensive glycan shield, surrounding the receptor-binding domain and can infect cells that express ABH antigens as a consequence of the individual phenotype~\cite{doi:10.1146/annurev.immunol.25.022106.141706, Watanabe2020}.
Importantly, these hypothesized asymmetrical transmission
rules should affect the number of people with a certain blood type that
can be potentially be infected because they do not allow all infected to infect any potentially contacted person. Consequently, we have
a dilution effect on the contagion that must be reflected in the growth of the epidemics. In the present work, we have studied how the existence of asymmetrical virus transmission rules affects such growth,
and how the fraction of infected patients with a certain blood type should not depend only on the patient blood type but on the number of infected people that may infect him.

To this aim, we expanded the usual SIR formalism to take into consideration the possible effect of blood antigenicity in the COVID-19 transmission.
Generally, to solve the set of generalized SIR Equations~(\ref{SIR-xg})-(\ref{SIR-zg}) for any time, researchers have relied on numerics. Here, we suggest a different approach, and we provided analytical solutions in the small-time limit, where the epidemics is in its exponential-growing regime. 
We obtained an expression linking the inverse characteristic time of the exponential phase with the abundances of the different blood groups in the population. We propose a set of susceptibility indices: $p_i^{(k)}$ for the sub-population $i$ and  $p_T^{(k)}$ for the total population. 

Here, we faced the problem from a general point of view and proposed a theoretical framework able to consider any set of rules based on blood groups. 
Then we focus on the case $k=2$ that describes a hypothetical set of rules based on the Rhesus group ($i$ may be 1 or 2 respectively for Rh$+$ and Rh$-$) and the case $k=4$ for the AB0 system.

Given a set of rules, the model predicts a linear dependence of the observed epidemic inverse timescale $m$ with the total susceptibility.
To test the model, we first compare its predictions with the experimental data provided in~\cite{zhao2020relationship, Li2020}, where the blood type of infected people of two Chinese regions was collected. Comparing the population frequencies of blood groups with those found in the infected sub-population, we verified that the proposed contagion scheme well describes the observed frequencies, since the difference observed between the ABO blood type population distribution and the ABO blood type infected people distribution supports the validity of the infection scheme proposed in \cite{breiman2020harnessing}. A similar approach has been used to test the model for the Denmark data~\cite{Barnkob2020}, characterized by an overall higher statistics, and again we found good accordance with the model.  

Next, we consider all data reported in Table~\ref{tab1}. Notably, the linear trend remains although data points are more scattered. We noted that most of the added sets of data were obtained considering only severely ill individuals and hospitalized individuals as a negative control. 
While our model only accounts for the effects of blood groups in the transmission of the virus, the ABH phenotype could in principle impact both the course of the COVID disease and the general health of individuals. Consequently, considering severely ill patients and hospitalized people as controls may introduce biases in both the $d_i$ and $f_i$ distributions, which reflect on a higher dispersion of the data in Figure~\ref{fig_all}.  

To further validate our model,  we analyzed the infection curves of a large set of countries worldwide, comparing the characteristic time of the infection outbreak with the prediction of our model based on the known blood group distribution for each country. 
Note that in this second case, we do not have the information about the blood groups of the infected population but only on the whole population. Thus, we can compare the growth rate in the short time, exponential, phase with the ``susceptibility" $p^{(k)}_T $ proposed by the model.  

We managed to collect data on the blood frequencies and the infections for 78 different countries belonging to four well distinct geographical areas, i.e. Europe, Asia, South America, and Africa. Since we expect that differences both geographical and on the lifestyles affect the infection and recovering rates we kept the countries separated according to the four identified areas.
This allows us to discard possible bias due to data collection and geographical factors.
Using a set of rules based on Rhesus blood types, we observed an always not significant correlation. From the point of view of the biological interpretation of the results, this is not surprising. A set of rules solely based on the Rh factor would not support the working hypothesis, which is based on the presence of glycan antigens on the viral capsid. 
 
On the contrary, considering the ABO set of rules, three out of the four areas present a very good agreement between data and model prediction (see Table~\ref{tab3}).  Summing up, it seems that the AB0 blood type has an effect on the virus spreading pattern.

In a nutshell, we proposed a generalized SIR model with infection rules dictated by antigenicity between different blood types. Obtaining an analytical solution of the model for the exponential phase, we were able to provide a rigorous theoretical test of the hypothesis proposed in~\cite{breiman2020harnessing}. We have made the test both for local data, where the number of infected people is stratified by blood type, and, on a wider scale, analyzing the infection growth curves of 78 countries worldwide. Overall, the present study reaches the conclusion that the hypothesis of a blood type effect on the COVID-19 advanced in ~\cite{breiman2020harnessing} it is not falsified by the available observational epidemic data. 
Obviously, to strengthen the validation of the hypothesis in  ~\cite{breiman2020harnessing} a direct detection of the antigens linked to the SARS-CoV-2 is needed, but this goes far beyond the goals of the present paper, which aims at presenting a mathematical framework to validate the hypothesis in  ~\cite{breiman2020harnessing} on patient level and country level data, and to allow us to understand the relation between the regional infection growth rate and the population blood type distribution. As a final note, we observe that, besides blood types, other population dependent antigens distribution may play a role in the geographically heterogeneous infection spreading~\cite{note}.



\section*{Additional information}
The authors declare that the research was conducted in the absence of any commercial or financial relationships that could be construed as a potential conflict of interest.

\bibliography{sample}

\begin{thebibliography}{54}
\bibitem[{not()}]{note}
\bibinfo{note}{As an example, it is worth to note that the distribution of the
  HLA types is highly variable in Europe. It is suggestive to consider that
  HLA-DRB1*11:01 shows a large peak in the northern Italian regions, where the
  epidemic has been by far more severe than in the rest of the continent.}
\end{thebibliography}

\end{document}